\documentclass[prb,prl,10pt,oneside,twocolumn]{revtex4}

\usepackage{graphicx}
\usepackage{bm}
\usepackage{gensymb}
\usepackage{amsmath}
\usepackage{epstopdf}
\usepackage{multirow}
\usepackage[sort&compress]{natbib}

\setcitestyle{square}

\begin{document}

\title{Non-Fermi Liquid Behavior Close to a Quantum Critical Point \\ in a Ferromagnetic State without Local Moments}

\author{E. Svanidze,$^1$ L. Liu,$^2$ B. Frandsen,$^2$ B. D. White,$^3$ T. Besara,$^4$ T. Goko,$^{2}$\footnote{Present Address: Laboratory for Muon Spin Spectroscopy, Paul Scherrer Institut, CH-5232 Villigen PSI, Switzerland} T. Medina,$^5$ T. J. S. Munsie,$^5$ G. M. Luke,$^5$ D. Zheng,$^6$ C. Q. Jin,$^6$ T. Siegrist,$^4$ M. B. Maple,$^3$ Y. J. Uemura,$^2$  and E. Morosan$^1$}

\address{$^1$Department of Physics and Astronomy, Rice University, Houston, TX 77005 USA}
\address{$^2$Department of Physics, Columbia University, New York, NY 10027 USA}
\address{$^3$Department of Physics, University of California, San Diego, La Jolla, CA 92093 USA}
\address{$^4$ National High Magnetic Field Laboratory, Florida State University, Tallahassee, FL, 32306 USA}
\address{$^5$Department of Physics and Astronomy, McMaster University, Hamilton, ON L8S 4M1, Canada}
\address{$^6$Institute of Physics, Chinese Academy of Sciences, Beijing 100190, China}

\date{\today}

\begin{abstract}

A quantum critical point (QCP) occurs upon chemical doping of the weak itinerant ferromagnet Sc$_{3.1}$In. Remarkable for a system with no local moments, the QCP is accompanied by non-Fermi liquid (NFL) behavior, manifested in the logarithmic divergence of the specific heat both in the ferro- \textsl{and} the paramagnetic states. Sc$_{3.1}$In displays critical scaling and NFL behavior in the ferromagnetic state, akin to what had been observed only in $f$-electron, local moment systems. With doping, critical scaling is observed close to the QCP, as the critical exponents $\delta$, $\gamma$ and $\beta$ have weak composition dependence, with $\delta$ nearly twice, and $\beta$ almost half of their respective mean-field values. The unusually large paramagnetic moment $\mu_{PM} \sim 1.3 \mu_B/F.U.$ is nearly composition-independent. Evidence for strong spin fluctuations, accompanying the QCP at $x_c = 0.035 \pm 0.005$, may be ascribed to the reduced dimensionality of Sc$_{3.1}$In, associated with the nearly one-dimensional Sc-In chains.

\end{abstract}


\maketitle

\section{Introduction}

Quantum critical points (QCPs) are ubiquitous features in the phase diagrams of strongly correlated electron systems, ranging from high temperature oxide superconductors \cite{ore, sac, boe} and low-dimensional compounds \cite{dai, gel, het}, to itinerant magnets (IMs) \cite{sok, sub, cad, yam} and heavy fermions (HFs) \cite{sch, tro, bud, don, ste, mori}. Often, non-Fermi liquid (NFL) behavior \cite{mil, tsv, geg, con, col}, and critical scaling \cite{kna} accompany the QCP, and such novel phenomena have been extensively studied in HFs. However, much less is currently understood about itinerant electron magnets and their quantum critical behavior, particularly due to the limited number of existent IMs. Of these, itinerant \textit{ferromagnets} (IFMs) are particularly appealing, since theoretical predictions suggest that the proximity to a ferromagnetic instability precludes the occurrence of a quantum phase transition (QPT). The QCPs recently observed in two substantively different systems, the IFM ZrZn$_2$ \cite{sok} and the HF ferromagnet URh$_2$Si$_2$ \cite{but}, are at odds with this prediction. Furthermore, NFL behavior is associated with the quantum phase transition induced by doping in the latter compound, but not the former, reemphasizing the imperious need for a unified picture of quantum criticality and NFL behavior in IFM systems. This study of the doping-induced NFL state close to the QCP in the IFM Sc$_{3.1}$In provides a first connection between the two previously known ferromagnetic QCP systems, a precursor of such a unified theory.

\begin{figure}
\includegraphics[width=1\columnwidth]{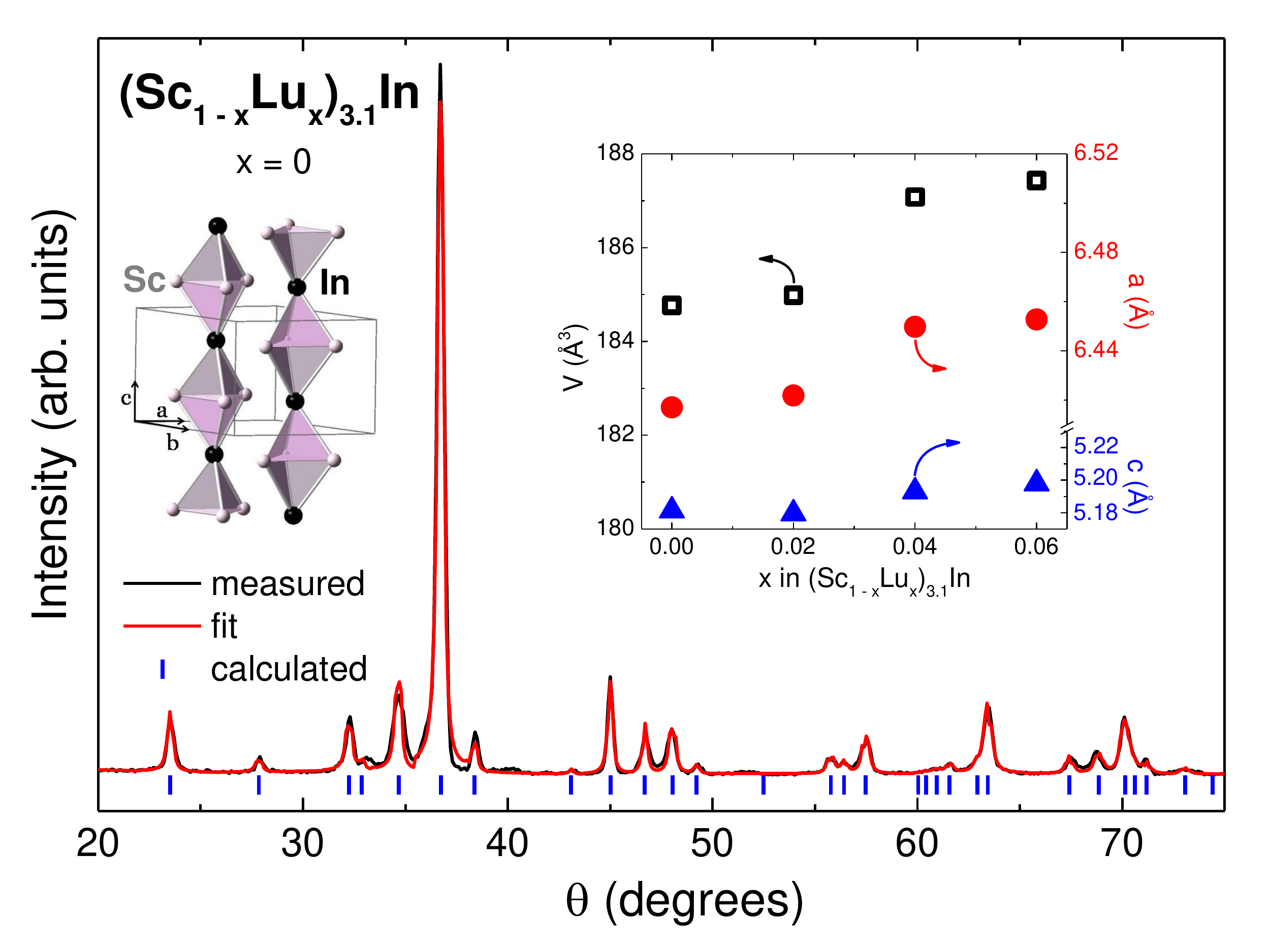}
\caption{Measured X-ray diffraction pattern for (Sc$_{1 - x}$Lu$_x)_{3.1}$In (black line) where $x = 0$, with calculated peak positions marked by blue vertical lines, based on space group P$6_3/mmc$ and lattice parameters $a = 6.42$ \AA~ and $c = 5.18$ \AA. The crystal structure of Sc$_{3.1}$In (left inset) exhibits quasi-1D chains of Sc-In. Right inset: evolution of lattice parameters $a$ and $c$ and the unit cell volume $V$ with composition $x$.}
\label{Xrays}
\end{figure}

Lu doping in Sc$_{3.1}$In represents the first report of NFL behavior associated with a QPT in this IFM. The critical composition $x_c$ in (Sc$_{1 - x}$Lu$_x)_{3.1}$In is very small, close to 0.035. The critical scaling close to the QCP is remarkable by comparison to ZrZn$_2$ or URu$_2$Si$_2$: while Sc$_{3.1}$In is similar to the former compound as the only other known FM with no magnetic elements, its critical scaling is not mean-field-like, akin to that in the latter system. The reduced dimensionality of Sc$_{3.1}$In, associated with quasi-1D Sc-In chains (inset of Fig. \ref{Xrays}), provides a plausible justification for the similarities with the two-dimensional URu$_2$Si$_2$, and contrasts with the three-dimensional ZrZn$_2$. In URu$_2$Si$_2$ the Kondo effect is inherently coupled with the quantum critical behavior, but Sc$_{3.1}$In has no local moments, rendering its magnetism and the QCP even more striking. What makes Sc$_{3.1}$In even more unique is the NFL behavior, a trait so far only present in $f$-electron systems. It is therefore paramount to probe the existence of the QCP in this IFM system, and properly characterize the NFL behavior, as a precursor for a unified picture of quantum criticality in IFMs. The development of such a unified theory necessitates more IFM systems, which starts with a thorough understanding of the few compounds that are already known.

IFMs lack the complexity associated with the interplay between the local and itinerant character of the electrons observed in HFs \cite{agu,ste}. Of the two known IFMs with no magnetic elements, ZrZn$_2$ and Sc$_3$In, the latter presents the advantage, from the quantum criticality perspective, of a much lower magnetic ordering temperature $T_C \leq 7.5$ K \cite{mat,gar,ike,tak,gre} in Sc$_3$In. This would likely facilitate the suppression of magnetic order towards a QCP, but has proven difficult by the application of pressure \cite{gre} or magnetic field \cite{ike}. Here we show that the QCP in Sc$_{3.1}$In can indeed be reached by Lu doping, where the dopant ion is comparable in size ($r[Lu^{3+}] = 0.861$ \AA) with the host ion Sc that it substitutes for ($r[Sc^{3+}] = 0.745$ \AA) \cite{sha}. This way, the effects of chemical substitution can be deconvoluted from those of chemical pressure, which is important given that pressure was shown to enhance the ordering temperature \cite{gre}. The systematic analysis of the magnetization isotherms $M(H)$, temperature-dependent magnetization $M(T)$ at low fields, and $H = 0$ specific heat data, for $0 \leq x \leq 0.10$ indicate that the magnetic ground state is suppressed in (Sc$_{1 - x}$Lu$_x)_{3.1}$In towards a QCP close to $x_c \approx 0.035$. Remarkably, the logarithmic divergence of the specific heat close to $x_c$ evidences NFL behavior in both the ferromagnetic and the paramagnetic state. Additionally, the reduced dimensionality, associated with quasi-1D Sc-In chains, may be linked to the NFL behavior and the non-mean-field critical scaling, similar to that in the more 2D FM, albeit with substantively different critical exponents.

\section{Experimental Methods}

The hexagonal Sc$_{3.1}$In compound has a Ni$_3$Sn-type structure, with space group P6$_3/mmc$ and lattice parameters $a = 6.42$ \AA~and $c = 5.18$ \AA~\cite{com}. The reported crystal structure for Sc$_3$In is shown in the left inset of Fig. \ref{Xrays}. Highlighted are the Sc-In bypiramids which form nearly one dimensional chains along the hexagonal $c$ axis. Band structure calculations \cite{jeo2} indicate that this crystal structure also renders the electronic configuration nearly one dimensional. These observations will be discussed in the context of the dimensionality of other IFM systems close to quantum criticality. 

It had already been established \cite{mat} that Sc$_{3}$In forms non-stoichiometrically around the ionic ratio Sc:In = 3:1. In the current study, we have determined that the optimal composition, which yielded the highest Curie temperature $T_C$ and paramagnetic moment $\mu_{PM}$, was Sc:In = 3.1:1. Polycrystalline samples of (Sc$_{1 - x}$Lu$_x)_{3.1}$In ($0 \leq x \leq 0.10$) were prepared by arcmelting Sc (Ames Laboratory, 99.999\%), Lu (Ames Laboratory, 99.999\%) and In (Alfa Aesar, 99.9995\%), with mass losses no more than $0.5\%$. The arcmelted buttons were subsequently wrapped in Ta foil, sealed in quartz tubes under partial Ar atmosphere, and annealed over two weeks at temperatures between $700^{\circ}$C and $800^{\circ}$C. 

Both annealed and non-annealed samples exhibit extreme hardness, comparable to that of high carbon steels \cite{sva}, which made it very difficult to perform powder x-ray diffraction measurements. However, it was feasible to x-ray a polished surface of the arcmelted buttons. The arcmelted samples with radius of about 3 mm were cut, and the flat surface was scanned for 12 hours in a Rigaku D/Max diffractometer with CuK$\alpha$ radiation and a graphite monochromator. An example of a diffraction pattern is shown in Fig. \ref{Xrays} for (Sc$_{1 - x}$Lu$_x)_{3.1}$In with $x = 0$. All observed peaks can be indexed with the space group P6$_3/mmc$. As shown in the right inset of Fig. \ref{Xrays}, both $a$ (right axis, circles) and $c$ (right axis, triangles) lattice parameters, along with the unit cell volume $V$ (left axis, squares) for (Sc$_{1 - x}$Lu$_x)_{3.1}$In for $0~\leq~x~\leq ~0.10$, increase nearly linearly with $x$. 

DC magnetization measurements on the annealed samples were performed in a Quantum Design (QD) Magnetic Property Measurement System for temperatures between $1.8$ K and $300$ K, and for applied magnetic fields up to 5.5 T. Specific heat was measured from $0.4$ K to $20$ K in a QD Physical Property Measurement System.

Measurements of AC magnetic susceptibility were performed in a $^4$He dewar down to $\sim$1.17 K, with temperatures below 4.2 K achieved by pumping on the He$^4$ bath with a Stokes pump. The AC magnetic susceptibility coils were positioned in the thermal gradient above the $^4$He bath by manually adjusting the vertical position of the probe.  An AC current was driven on the primary coils with a frequency of 15.9 Hz using a Linear Research LR700 AC resistance bridge, which produces an AC magnetic field with magnitude of $\sim 0.3$ Oe. This bridge was also used to measure the in- and out-of-phase components of the signal induced in the secondary pickup coils. The secondary coils are balanced by counter-winding the wire to cancel background signals induced by the oscillating AC magnetic field. A small offset in the measured signal due to minor imbalances in the home-built AC susceptibility coils was subtracted from the data. The data were then scaled so that their arbitrary units are proportional to emu/mol.

\begin{figure*}
\includegraphics[width=2\columnwidth]{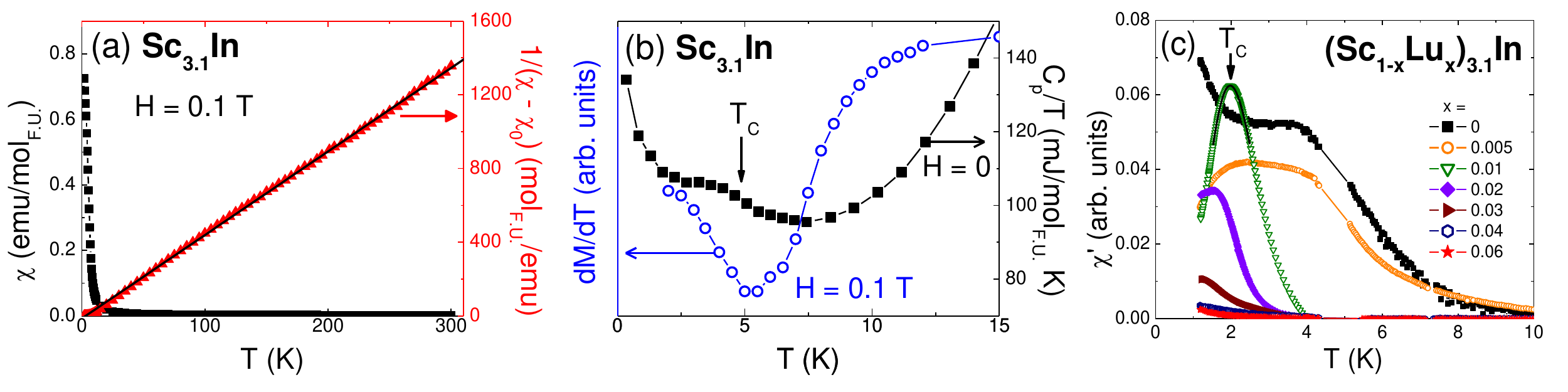}
\caption{(a) Sc$_{3.1}$In susceptibility (left) and inverse susceptibility $1/(\chi - \chi_0)$ (right) for $H = 0.1$ T, where $\chi_0 = C/T^*$ (see text). (b) The magnetization derivative $dM/dT$ (left) and specific heat $C_p/T$ (right) for Sc$_{3.1}$In with the vertical arrow marking the Curie temperature $T_C$. (c) (Sc$_{1 - x}$Lu$_x)_{3.1}$In AC susceptibility  $\chi'$(T). The Curie temperature $T_C$ is estimated from the peak position (solid line), indicated by the vertical arrow.}
\label{MT}
\end{figure*}

Muon Spin Relaxation ($\mu$SR) measurements were performed at TRIUMF using a He gas flow cryostat at the M20 beamline for Sc$_{3.1}$In and another spectrometer with dilution cryostat at the M15 beamline for $x = 0.01$ and $x = 0.025$ (Sc$_{1-x}$Lu$_{x})_{3.1}$In samples. Details of the $\mu$SR technique can be found elsewhere \cite{Schenck_1985, Lee_1999, Heffner_2004, Yaouanc_2010, Sonier_2002}.  

\section{Results and Analysis}

\subsection{Temperature-Dependent Magnetization Measurements}

For weak IFMs, the low-field susceptibility is expected to follow a $T^{-1}$ behavior \cite{mor}: 

\begin{equation}
\frac{\chi_0}{\chi(T)} = 1 - \alpha + \lambda(T), 
\end{equation}

\noindent where the coefficient $\lambda$ encompasses the dependence on the local amplitude of spin fluctuations and is linear in temperature $\lambda \sim T/T^*_C$, and $\alpha = I \rho(E_F)$, where $I$ is the coupling constant and $\rho(E_F)$ is the density of states at the Fermi level. When $T^* \gg T^*_C$, $\alpha \sim (T/T^*)^2$ usually has only a weak $T^2$ dependence. However, the magnetic susceptibility $\chi (T)$ in Sc$_{3.1}$In follows a modified Curie-Weiss-\textit{like} law: 

\begin{equation}
\chi(T) = \frac{C}{T^*} + \frac{C}{(T - T^*_C)}, 
\end{equation}

\noindent as illustrated in Fig. \ref{MT}(a). Such a temperature dependence can possibly be understood when considering strong spin fluctuations, associated with the nearly one-dimensional Fermi surface of Sc$_3$In \cite{jeo2}: if $T^* > T^*_C$ (but \textit{not} $\gg T^*_C$), then the temperature dependence of $\alpha$ is not negligible compared to that of $\lambda$. In weak IFMs, the Curie-Weiss-\textit{like} temperature $T^*_C$, determined from linear fits of the inverse susceptibility after the temperature independent term $C/T^*$ was subtracted, coincides with the Curie temperature $T_C$. As shown below, this is not the case in (Sc$_{1 - x}$Lu$_x)_{3.1}$In, even though $T_C$ and $T^*_C$ are both continuously suppressed to 0 K with $x$.

A local minimum in the derivative $dM/dT$ (Fig. \ref{MT}(b), left axis) corresponds to the Curie temperature $T_C$ in Sc$_{3.1}$In. Moreover, the specific heat data for $x = 0$, plotted as $C_p/T$ (right axis, Fig. \ref{MT}(b)), also displays a broad maximum at $T_C$. This is remarkable, given that such transitions are often difficult to identify in the field-independent properties of IFMs, even in single crystalline samples \cite{but}. In Sc$_{3.1}$In, the susceptibility derivatives and specific heat data provide evidence that the ferromagnetic ordering occurs below $T_C \sim 4.5$ K, as also demonstrated by the field-dependent data shown below. The different measurements consistently indicate that $T_C$ is significantly lower than the older estimates from Arrott isotherms alone \cite{gar,tak,gre}, when Sc$_3$In was erroneously assumed to be a mean-field ferromagnet.

In (Sc$_{1 - x}$Lu$_x)_{3.1}$In, $T_C$ is continuously suppressed by Lu doping above $x = 0.02$ to values below those accessible by the QD MPMS system. Further data below $T = 2$ K was collected from $^4$He AC susceptibility measurements shown in Fig. \ref{MT}(c). Lack of data around the 4.2 K $^4$He transition precludes a $T_C$ estimate for $x = 0$ and $x = 0.005$, when the transition falls close to this temperature interval. However, for all other compositions up to $x = 0.04$, the peak corresponding to $T_C$ (illustrated by the solid line fit in Fig. \ref{MT}(c)) is continuously reduced to temperatures below $T = 1.17$ K, as shown in Fig. \ref{MT}(c). This agrees with the critical composition $x_c = 0.035 \pm 0.005$, as determined from the analysis below.

\begin{figure*}
\includegraphics[width=2\columnwidth]{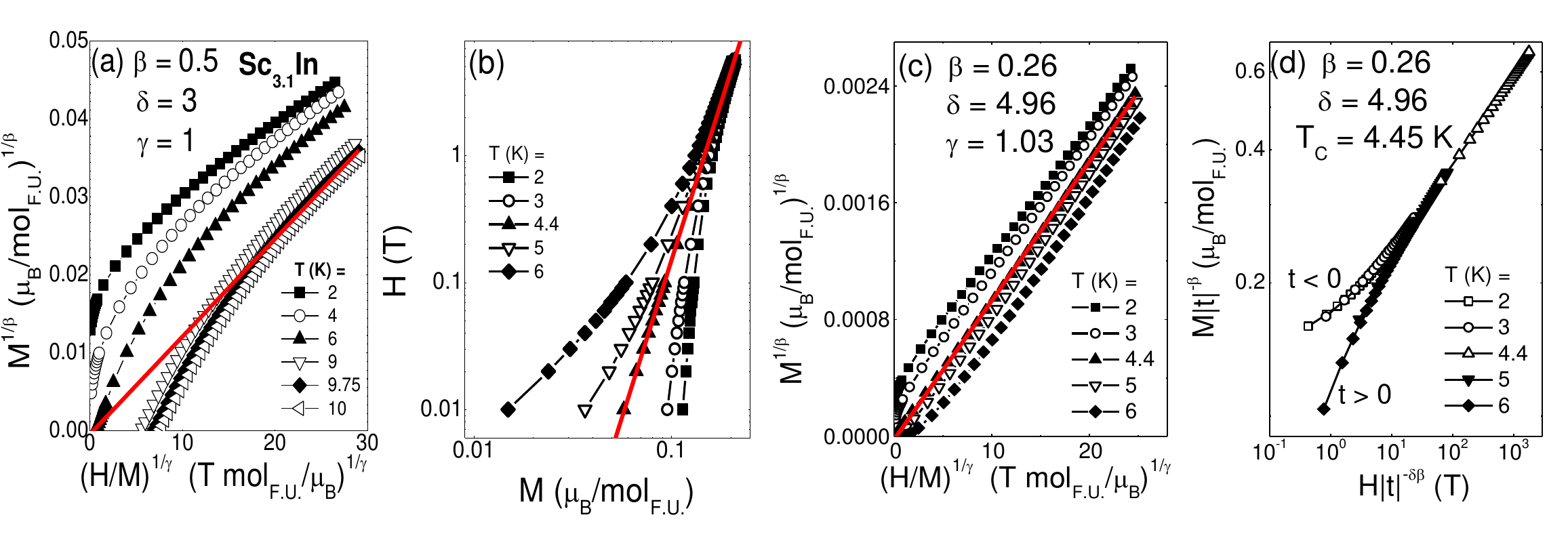}
\caption{$M^{1/\beta}$ \textit{vs.} $(H/M)^{1/\gamma}$ isotherms for Sc$_{3.1}$In with (a) mean field exponents $\beta = \beta_{MF} = 0.5$ and $\gamma = \gamma_{MF} = 1$, $T_C = 9.75$ K (solid line) and (c) non-mean-field exponents $\beta = 0.26$ and $\gamma = 1.03$, $T_C = 4.45$ K (solid line). (b) Log-log plot of Sc$_{3.1}$In M(H) isotherms, with the straight line representing the fit for the critical isotherm. (d) Arrott-Noakes scaling plots $M|t|^{-\beta}$ \textit{vs.} $H|t|^{-\delta \beta}$. The scaled $M(H)$ data collapses onto two diverging branches, one below $T_C$ ($t < 0$, open symbols) and one above $T_C$ ($t > 0$, full symbols).}
\label{Sc3In}
\end{figure*}

\begin{figure*}
\includegraphics[width=2\columnwidth]{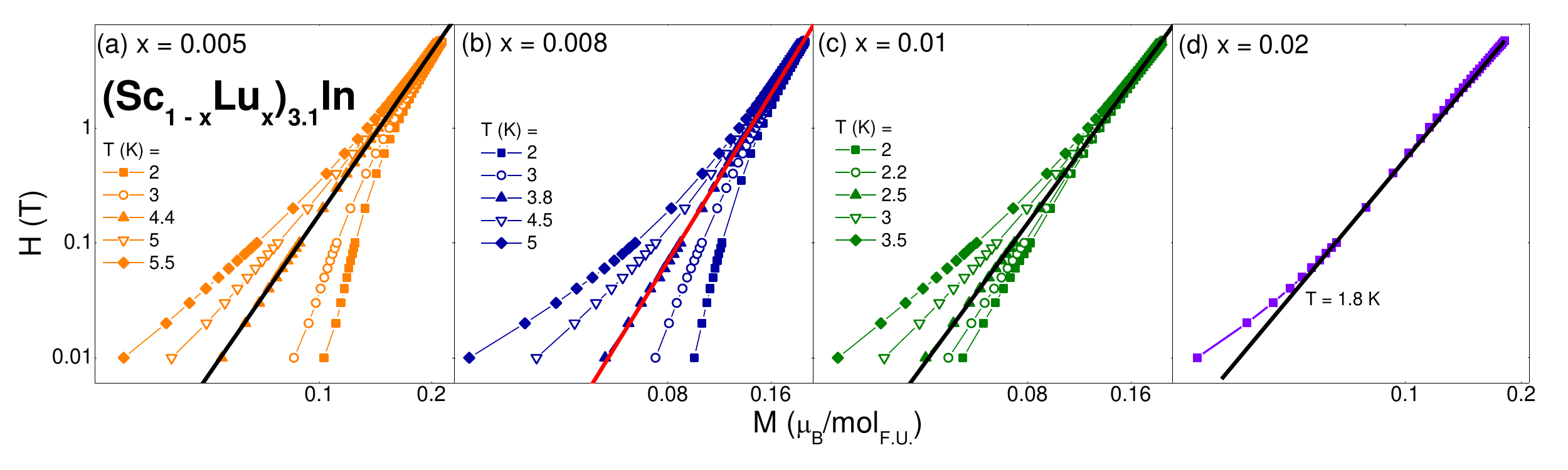}
\caption{Log-log (Sc$_{1-x}$Lu$_x$)$_{3.1}$In M(H) isotherms for (a) $x = 0.005$, (b) $x = 0.008$ and (c) $x = 0.01$ with linear fits (solid lines) at the critical ($T = T_C$) isotherm. (d) Log-log $M(H)$ curve for $x = 0.02$ and $T = 1.8$ K, with a linear fit above $H = 0.05$ T.}
\label{LogMH}
\end{figure*}

\subsection{Arrott and Arrott-Noakes Analysis}

Strong spin fluctuations in Sc$_{3.1}$In result in deviations from linearity in the inverse susceptibility around $T^*_C$ which precludes the accurate determination of the Curie temperature $T_C$ from the $\chi(T)$ data. Alternatively, Arrott isotherms $M^2$ \textsl{vs.} $H/M$ \cite{arr} had previously been employed to determine $T_C$ in Sc$_3$In. Existing reports give this value to be less than 7.5 K \cite{mat,gar,ike,tak,gre}. If the Arrott plot technique were used for Sc$_{3.1}$In (Fig. \ref{Sc3In}(a)), it would appear that ferromagnetic order occurred close to 9.75 K. This implies that the Sc-In ratio used for the current study is closest to the optimal one \cite{mat}, compared to all previous reports. However, the Arrott isotherms deviate strongly from linearity at high $H$ values \cite{gre,gar,tak}. This is a compelling indication that the mean-field theory cannot accurately describe the weak ferromagnetism in Sc$_{3.1}$In, in contrast with, for example, ZrZn$_2$ \cite{sok}. The more generalized Arrott and Noakes method\cite{noa} was successfully employed to characterize the critical scaling in the HF ferromagnet URu$_2$Si$_2$ doped by Re \cite{but}. In the current work, this generalized critical scaling is applied to a different type of QCP, in the weak IFM Sc$_{3.1}$In which has no local moment elements. It would appear that the NFL behavior results from the non-mean field character of the magnetism in these IFMs.

The Arrott-Noakes scaling represents a generalization of the mean-field scaling of the magnetization $M$, magnetic field $H$ and the reduced temperature $t = (T/T_C - 1)$:

\begin{equation}
M \propto t^{\beta} \hspace{5pt} \text{for} \hspace{5pt} t < 0
\end{equation}

\begin{equation}
M \propto H^{1/\delta} \hspace{5pt} \text{at} \hspace{5pt} t = 0
\end{equation}

\begin{equation}
\chi \propto t^{-\gamma} \hspace{5pt} \text{for} \hspace{5pt} t > 0
\end{equation}

\begin{figure*}
\includegraphics[width=2\columnwidth]{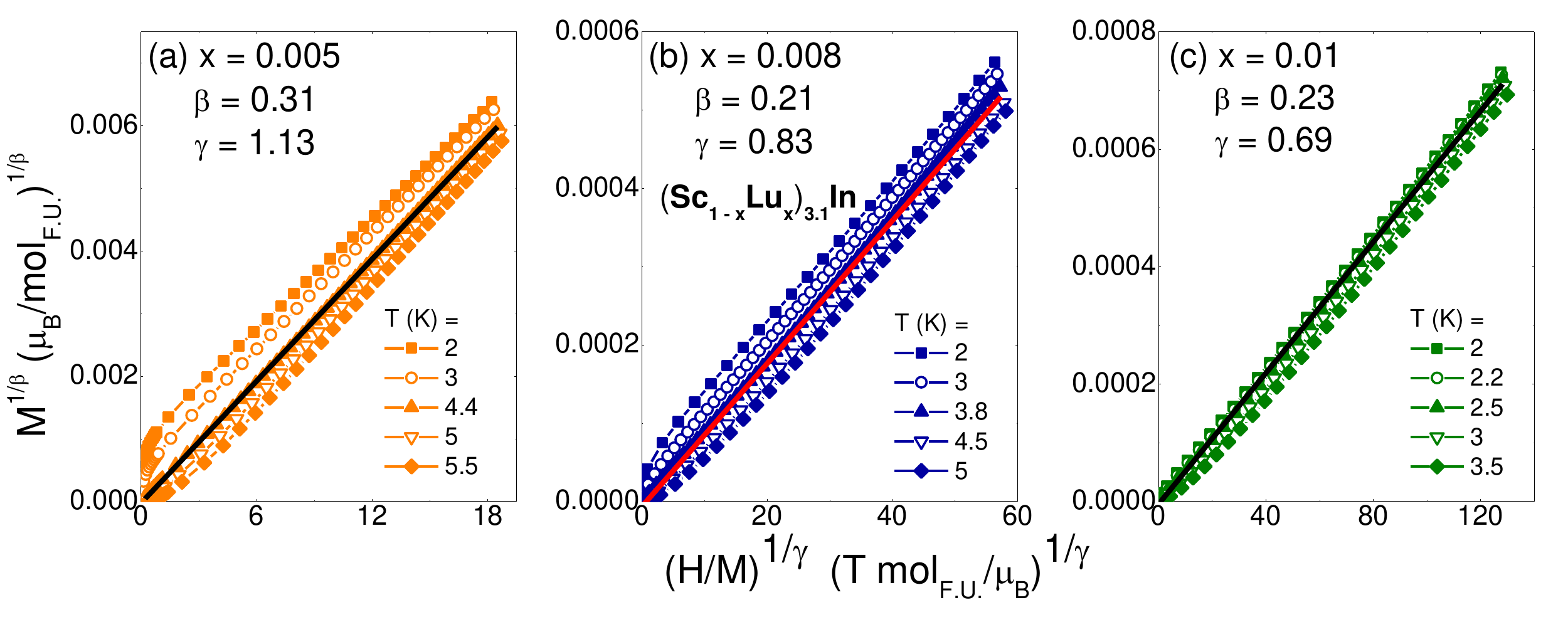}
\caption{(Sc$_{1 - x}$Lu$_x)_{3.1}$In Arrott-Noakes $M^{1/\beta}$ \textit{vs.} $(H/M)^{1/\gamma}$ isotherms for (a) $x = 0.005$, (b) $x = 0.008$, and (c) $x = 0.01$ with linear fits (solid lines) at $T_C$. }
\label{AN}
\end{figure*}

\noindent This yields generalized critical exponents $\beta$, $\delta$ and $\gamma$ with the constraint that $1 + \gamma/\beta - \delta = 0$. In the case of (Sc$_{1 - x}$Lu$_x)_{3.1}$In, the Curie temperature $T_C$ and exponent $\delta$ are first determined from log-log $M(H)$ plots for each composition, as shown in Fig. \ref{Sc3In}(b) for $x = 0$ and in Fig. \ref{LogMH}(a-c) for $x = 0.005, 0.008$, and $x = 0.01$. At $T_C$, critical scaling requires that the isotherm be linear, with a slope equal to the critical exponent $\delta$. The $T = 1.8$ K isotherm for $x = 0.02$ is nearly linear all the way down to $H = 0$ T (Fig. \ref{LogMH}(d)), indicating that $T_C$ for $x = 0.02$ is finite and smaller than 1.8 K. For all other FM compositions (Fig. \ref{LogMH}(a-c)), non-linear isotherms occur within 20$\%$ of $T_C$. Therefore, in the absence of $M(H)$ measurements below 1.8 K, the nearly linear log-log $M(H;1.8~\text{K})$ isotherm for $x = 0.02$ is a good indication that the $T_C$ value estimate for this composition is within 20$\%$ of $T_C$, which yields $T_C(x = 0.02) = 1.5 \pm 0.3$ K. This value agrees well with the AC susceptibility estimate, where $T_C = 1.62$ K.

\begin{figure}
\includegraphics[width=\columnwidth]{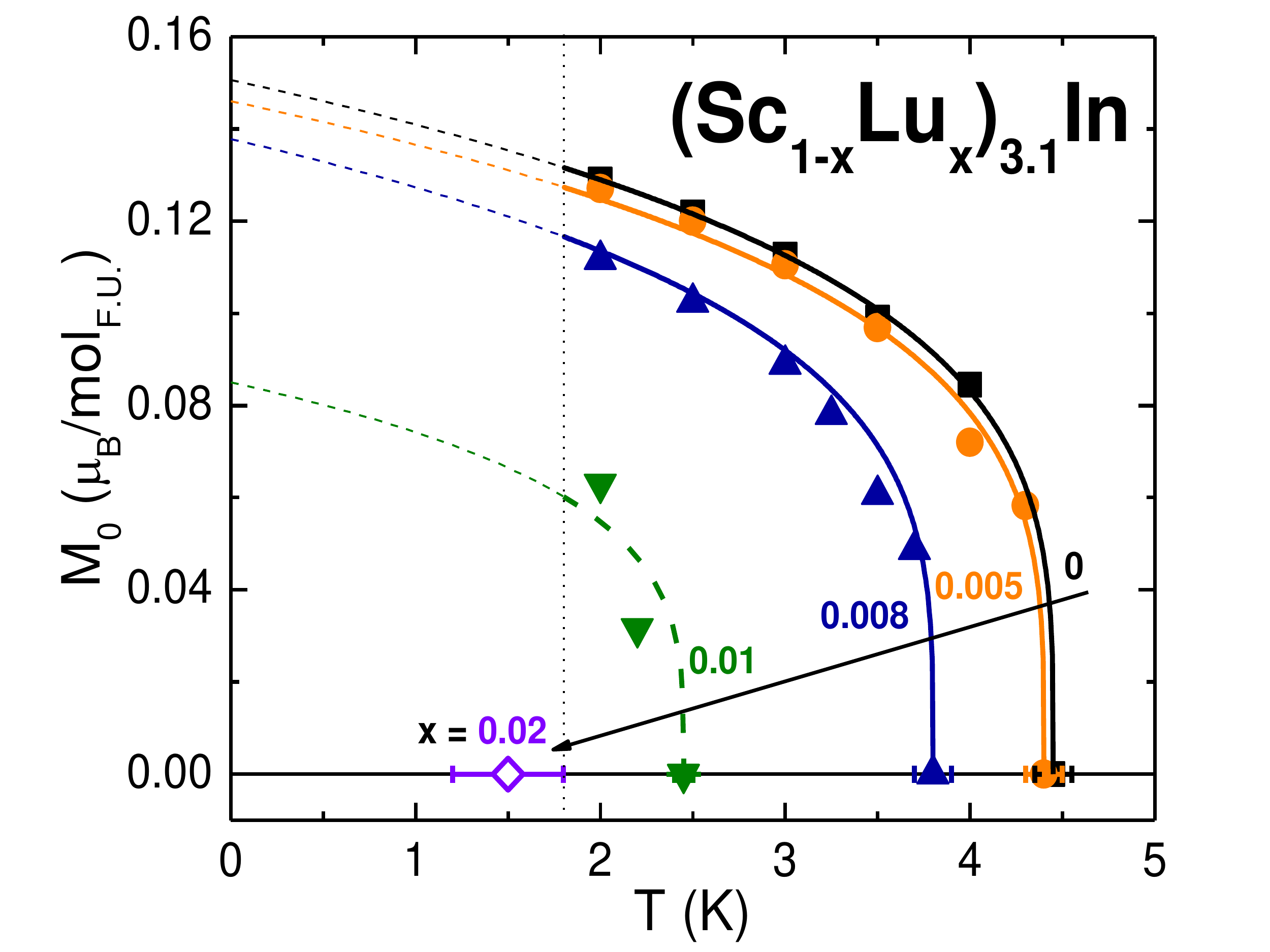}
\caption{Spontaneous magnetization $M_0(T)$ for (Sc$_{1 - x}$Lu$_x)_{3.1}$In where $0 \leq x \leq 0.02$.}
\label{M0}
\end{figure}

\begin{figure}
\includegraphics[width=\columnwidth]{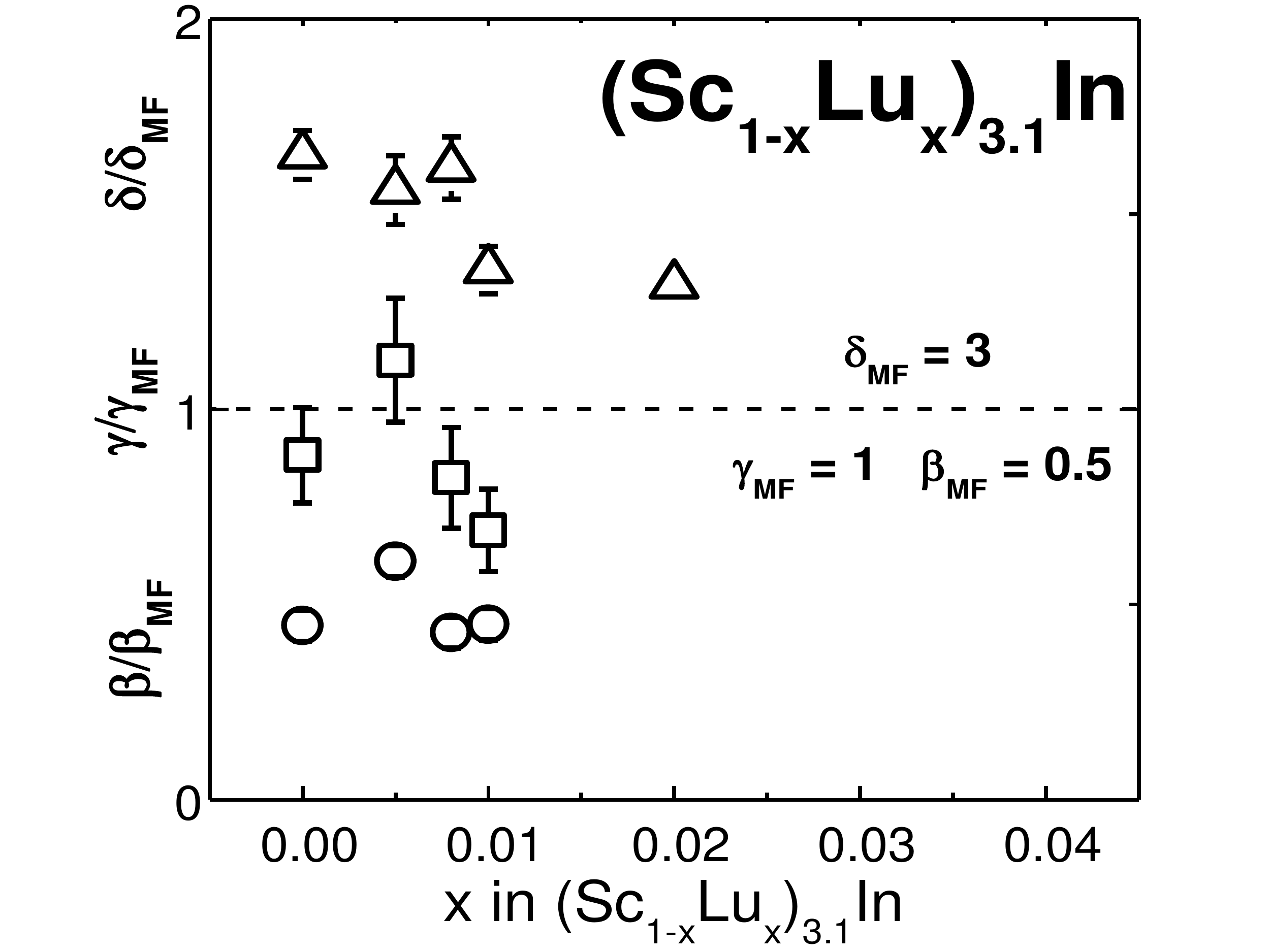}
\caption{The critical exponents scaled by their mean-field values $\delta/\delta_{MF}$ (triangles), $\gamma/\gamma_{MF}$ (squares) and $\beta/\beta_{MF}$ (circles) as a function of composition $x$.}
\label{Exponents}
\end{figure}

Next, the critical exponents $\beta$ and $\gamma$ are determined from the expected linear dependence of $M^{1/\beta}$ \textsl{vs.} $(H/M)^{1/\gamma}$. A subset of the resulting isotherms is shown in Figs. \ref{Sc3In}(c) ($x = 0$) and \ref{AN}(a-c) ($x = 0.005, 0.008$ and 0.01). The extrapolations of the linearized isotherms in the ferromagnetic state yield the spontaneous magnetization $M_0$ from the vertical axes intercepts. As expected, $M_0$ scales with $|t|^\beta$, as shown in Fig. \ref{M0} for (Sc$_{1 - x}$Lu$_x)_{3.1}$In where $0 \leq x \leq 0.02$. In contrast with URu$_{2-x}$Re$_x$Si$_2$ \cite{but}, $M_0$ for (Sc$_{1-x}$Lu$_x$)$_{3.1}$In (Fig. \ref{M0}) grows faster in the ordered state, as the critical exponent $\beta$ for the former, $\beta = 0.26 \pm 0.05$, is less than half of the respective value in the latter system \cite{but}. However, the $\beta$ values in (Sc$_{1-x}$Lu$_x$)$_{3.1}$In are unusually small, which implies that the ordered moment in this weak IFM is more readily destabilized by fluctuations close to $T_C$. This might indicate a fragile magnetism in a nearly 1D electron system \cite{jeo2}, which doping and the attendant disorder immediately suppress to 0 K.

\begin{figure*}
\includegraphics[width=2\columnwidth]{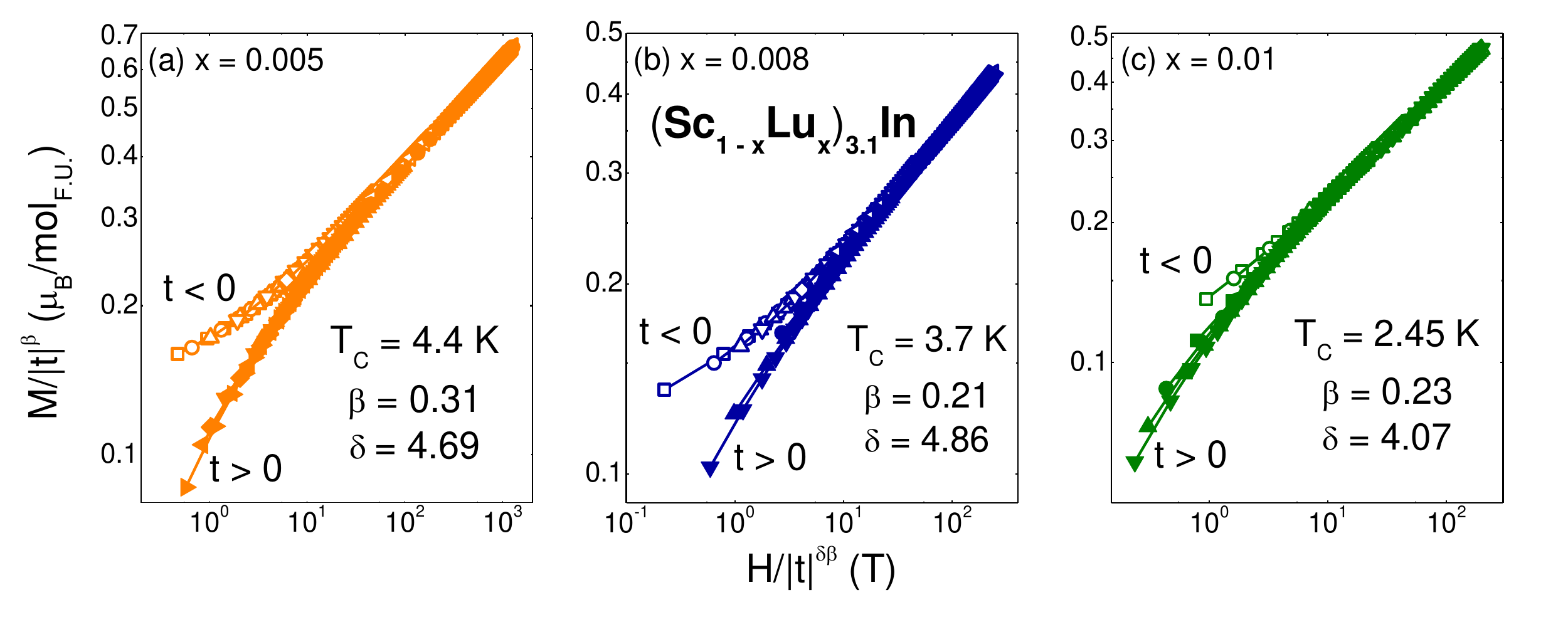}
\caption{Scaling plots $M|t|^{-\beta}$ \text{vs.} $H|t|^{-\delta \beta}$ for (a) $x = 0.005$, (b) $x = 0.08$ and (c) $x = 0.01$ in (Sc$_{1-x}$Lu$_x$)$_{3.1}$In.}
\label{Collapse}
\end{figure*}

The Arrott-Noakes critical exponents $\delta$ (triangles), $\gamma$ (squares) and $\beta$ (circles), scaled by their mean-field (MF) values, are presented in Fig. \ref{Exponents} as a function of composition. Most strikingly, $\delta$ is nearly twice as large as its mean-field value $\delta_{MF}$, while $\beta$ is nearly half of $\beta_{MF}$, leaving $\gamma$ nearly identical to its mean-field value $\gamma_{MF}$. $\delta$ is a measure of the curvature of $M(H)$ at $T_C$, with larger values signaling faster saturation. A comparison between (Sc$_{1-x}$Lu$_x$)$_{3.1}$In and URu$_{2-x}$Re$_x$Si$_2$ \cite{but} shows that larger $\delta$ values for the former compound are also associated with a larger relative magnetization $M(5.5~\text{T};1.8~\text{K}) \approx 0.2~\mu_B$. This value at $t = 0.6$ is nearly $15 \%$ of the paramagnetic moment $\mu_{PM} \approx 1.3~\mu_B$ for the composition $x = 0$ with maximum $T_C$. The corresponding value for URu$_{2-x}$Re$_x$Si$_2$ is $M(5.5~\text{T};1.8~\text{K})/ \mu_{PM} \approx (0.4~\mu_B)/(3.8~\mu_B) \approx$ 10$\%$ (for which $T_{C,max} = 27$ K is obtained for $x = 0.6$), nearly one third less at a comparable relative temperature $t$ (Fig. 1, bottom panel, in Ref. [24]). 

The scaling collapse of $M|t|^{-\beta}$ \textit{vs.} $H|t|^{-\delta \beta}$, shown in Figs. \ref{Sc3In}(d) and \ref{Collapse}(a-c), exemplifies how the $M|t|^{-\beta}$ \textit{vs.} $H|t|^{-\delta \beta}$ curves collapse onto two diverging branches, for $t~<~0$ (open symbols) and $t~>~0$ (full symbols). This collapse is similar to that observed for the HF URu$_{2-x}$Re$_x$Si$_2$ \cite{but}, which is remarkable, given the lack of formal local moments in the constituent elements of (Sc$_{1-x}$Lu$_x$)$_{3.1}$In.

\subsection{Non-Fermi Liquid Behavior}
\begin{figure}
\includegraphics[width=\columnwidth]{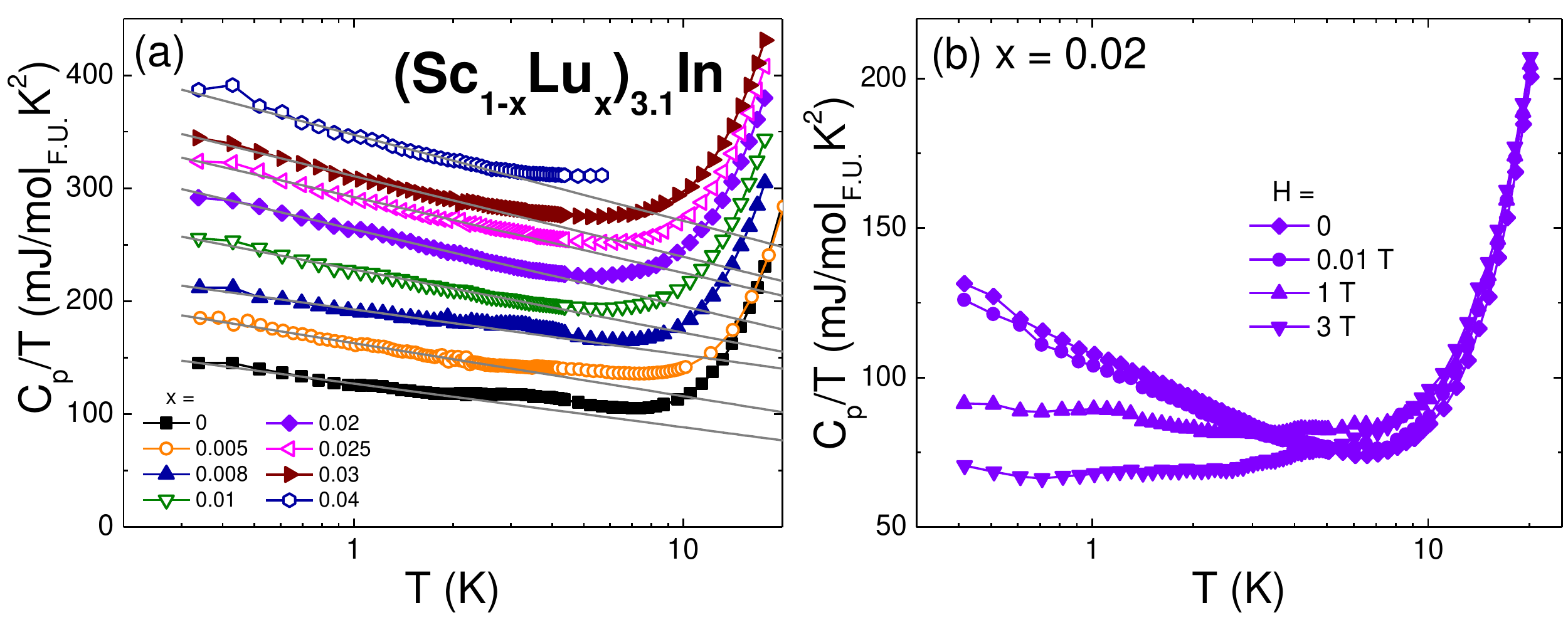}
\caption{(a) Semi-log plot of the specific heat $C_p/T$ (symbols) and the linear fits (lines) at low temperatures. The data for $x > 0$ are shifted vertically for clarity by 30 mJ/mol$_{F.U.}$ K$^2$. (b) Specific heat for (Sc$_{0.98}$Lu$_{0.02}$)$_{3.1}$In in various magnetic fields: $H$ = 0, 0.01, 1 and 3 T.}
\label{CpT}
\end{figure} 

An independent and compelling evidence for the QCP in the doped Sc$_{3.1}$In system is the NFL behavior below $x = 0.04$. The signature of NFL behavior is the logarithmic divergence of the specific heat $C_p/T$ (Fig. \ref{CpT}(a)), which occurs over a decade in temperature. This divergence of the specific heat, shown in Fig. \ref{CpT}(a), may have two possible origins: NFL behavior or Schottky anomaly. For a Schottky anomaly, a low-$T$ peak in the specific heat would move up in temperature with increasing $H$. However, the decrease of the low temperature $C_p/T$ with increasing $H$ (Fig. \ref{CpT}(b)) invalidates the Schottky anomaly scenario and not surprisingly, since this would be associated with low-lying energy states (not the case for a system with no formal local moments). The NFL scenario is therefore more plausible in (Sc$_{1 - x}$Lu$_x)_{3.1}$In for $0 \leq x \leq 0.04$. More interestingly, the NFL behavior coexists with the ferromagnetic state. This coexistence has been explained based on magnetic cluster formation as a result of competition between Ruderman-Kittel-Kasuya-Yosida (RKKY) coupling and Kondo effect \cite{bau, net}. However, this is the \textit{first} observation of NFL behavior within the ferromagnetic state in a weak IFM. The implication is that a new model would be required to describe the ground state in Sc$_{3.1}$In, or that the Griffiths-McCoy model \cite{gri} may still be appropriate if evidence for Kondo effect emerged for this compound.

\subsection{Muon Spin Relaxation Measurements}

\begin{figure}
\includegraphics[width=\columnwidth]{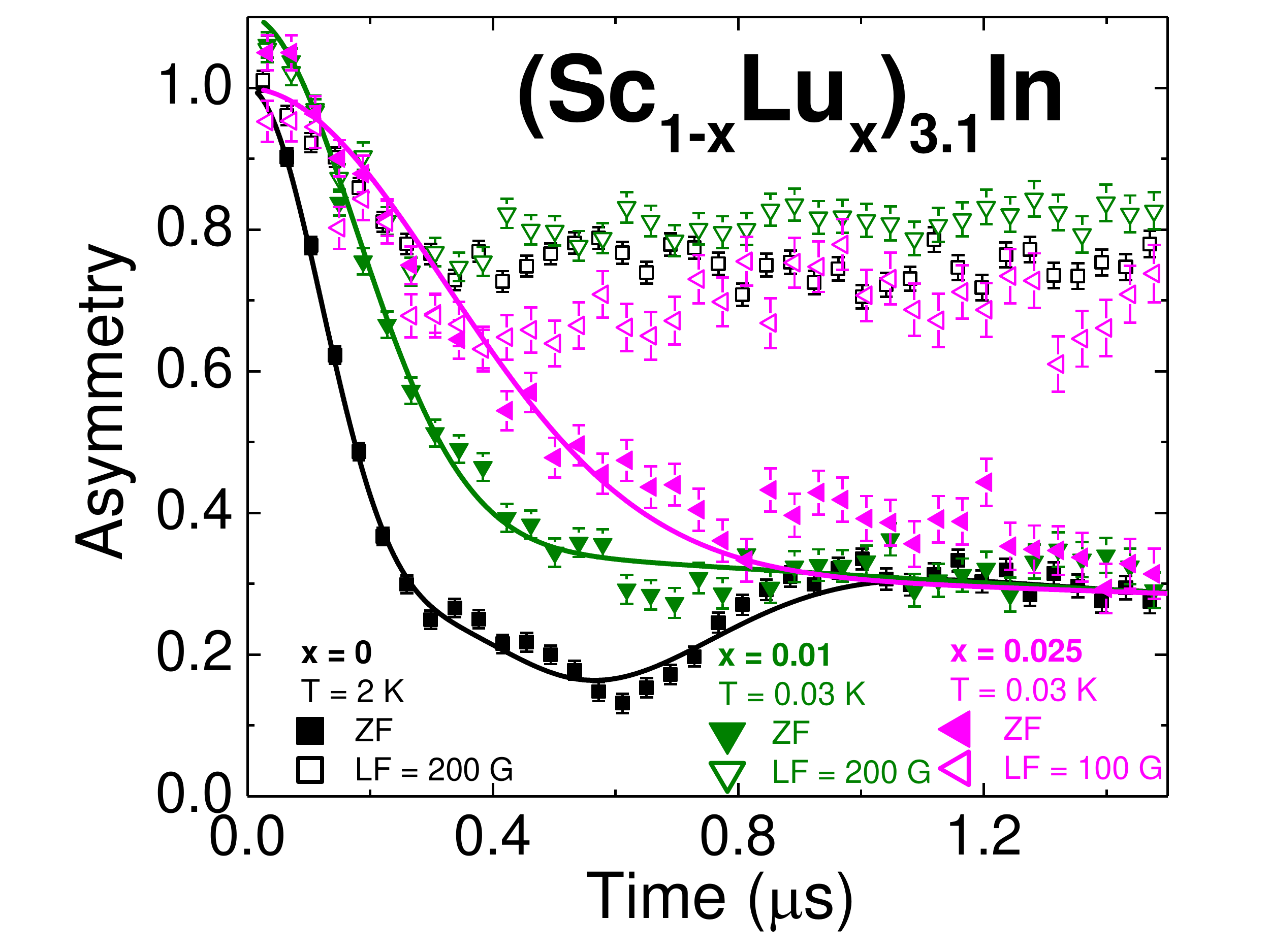}
\caption{Time spectra of ZF and LF $\mu$SR of (Sc$_{1-x}$Lu$_x$)$_{3.1}$In  where $x$ = 0 (squares), 0.01 (downward-facing triangles) and 0.025 (leftward-facing triangles). The background sample holder contribution in the two Lu-doped samples was subtracted. The solid lines represent fits to Eqs. \ref{Gt} and \ref{GtF}.}
\label{muSR1}
\end{figure}

\begin{figure}
\includegraphics[width=\columnwidth]{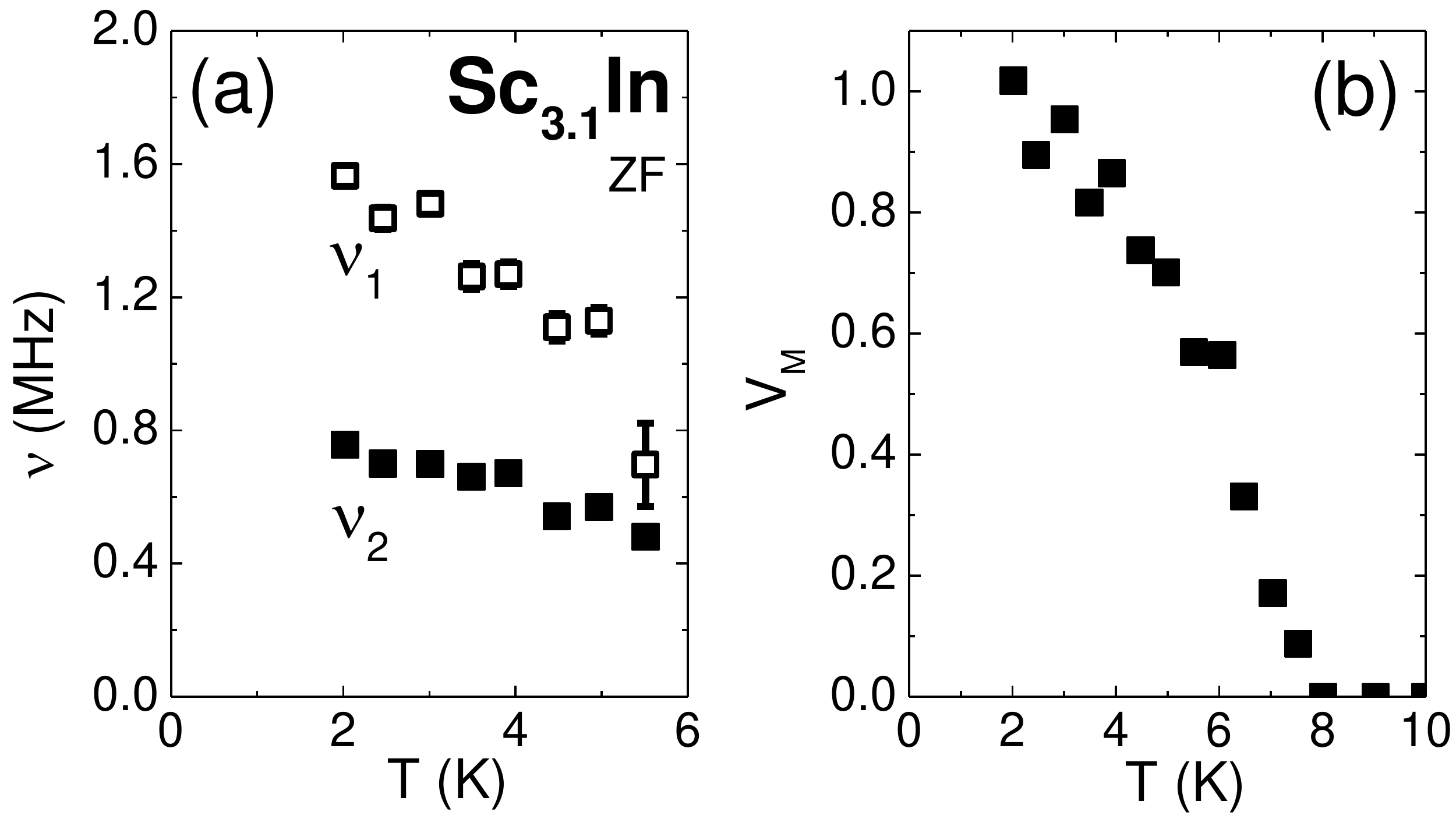}
\caption{(a) The muon spin precession frequencies $\nu_1$ (full squares) and $\nu_2$ (open squares), and (b) the volume fraction $V_M$ of the magnetically ordered regions, obtained from ZF $\mu$SR of Sc$_{3.1}$In}
\label{muSR2}
\end{figure}

The small saturation moments of itinerant systems preclude neutron diffraction investigations, similar to the case of ZrZn$_2$ \cite{Abrahams_1959}. On the contrary, the muon spin relaxation ($\mu$SR) technique is extremely sensitive to local magnetic fields and has been used to investigate multiple itinerant systems \cite{Uemura_2007, Uemura_1985, Gat_2011, Carlo_2012, Nozaki_2013}. Figure \ref{muSR1} shows the time spectra observed in zero field (ZF) and longitudinal field (LF) at the lowest temperature. The fast relaxation in the early time region in ZF is eliminated by the decoupling effect in LF, which indicates that the observed relaxation is due to a static field, generated by the static magnetic order in both the undoped and doped systems. For the ZF Sc$_{3.1}$In spectrum, two precession frequencies at low temperatures can be seen, as shown in Fig. \ref{muSR2}(a). The ZF time spectra were analyzed by assuming a functional form of:

\begin{widetext}
\begin{equation} \begin{split}
G(t)=[A_1\cos(2\pi \nu_1 t)e^{(-\Lambda_1^2 \frac{t^2}{2})} + A_2\cos(2\pi \nu_2 t)e^{(-\Lambda_2^2 \frac{t^2}{2})} + (A_{1Z}+A_{2Z})e^{(-\frac{t}{T_1})}]V_M + \\ \frac{(1-V_M)[G_{KT}(t,\Delta_{KT1})+G_{KT}(t,\Delta_{KT2})]}{2}
\label{Gt}
\end{split} \end{equation}
\end{widetext}

\noindent where $G_{KT}(t)$ is the Kubo-Toyabe function \cite{Hayano_1979} for random nuclear dipolar fields, and $A_{1Z}$ and $A_{2Z}$ are assumed to be a half of $A_1$ and $A_2$, respectively, as expected for polycrystalline specimens. A very good fit was obtained by assuming $A_1$ = $A_2$, presumably due to two magnetically-nonequivalent muon sites populated with equal probabilities. For the longitudinal relaxation rate $1/T_1$, two values for the two different sites could not be resolved. So, one value of $T_1$ was used in the fit. The temperature dependence of the two frequencies $\nu_1$ and $\nu_2$ is shown in Fig. \ref{muSR2}(a). The volume fraction $V_M$ of the magnetically ordered region, shown in Fig. \ref{muSR2}(b), was determined from the amplitudes of the precession signals. The volume fraction $V_M$ decreases gradually with increasing temperature, indicating co-existence of volumes (or regions) with and without static magnetic order. Although the precession signal disappears around $T = 5.5$ K, a small $V_M$ remains above this temperature up to $T \approx 8$ K. This is due to a non-precessing but relaxing signal with a small amplitude, caused by static random fields from the electron system remaining in a small volume fraction.

\begin{figure}
\includegraphics[width=\columnwidth]{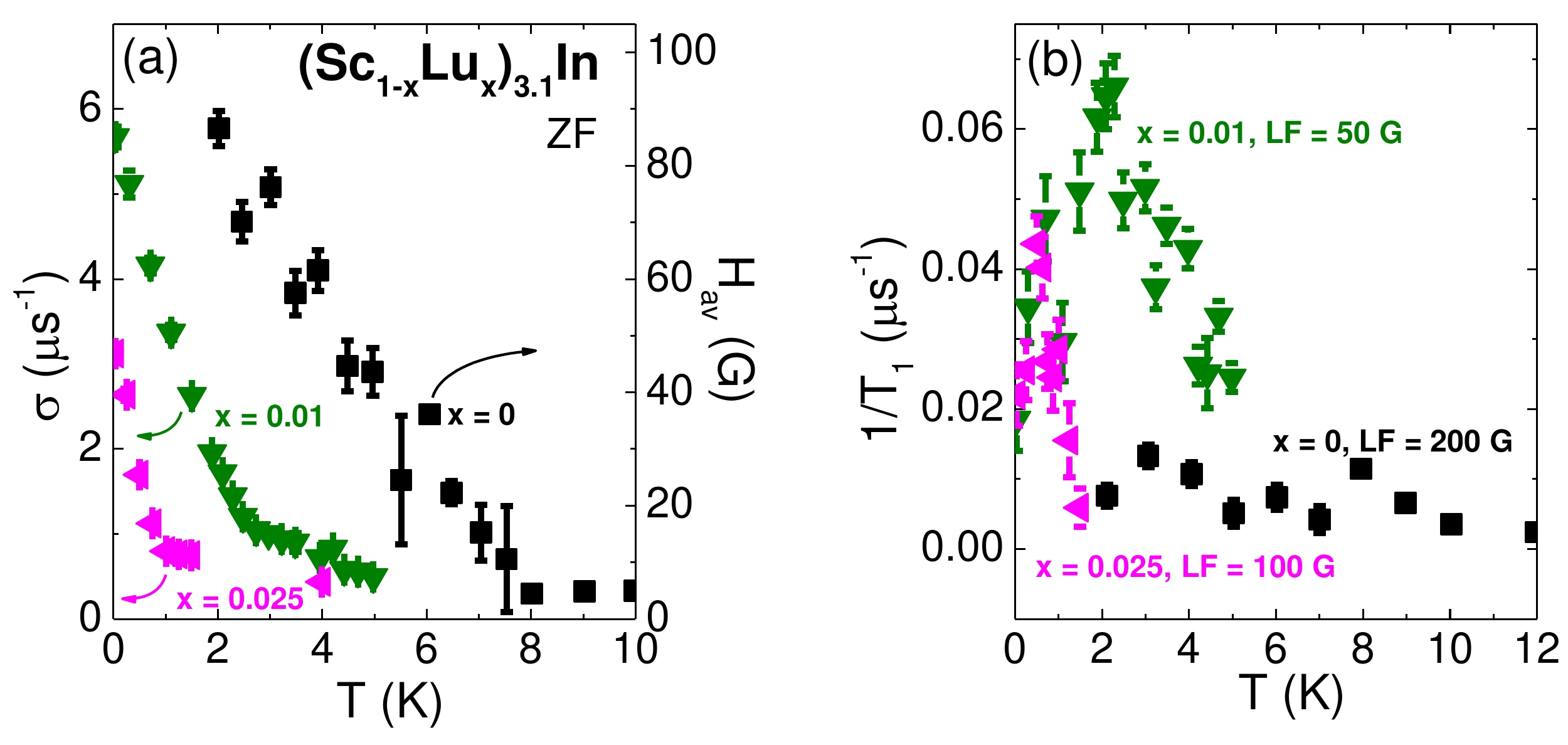}
\caption{(a) Muon spin relaxation rate $\sigma$ for $x = 0.01$ (downward-facing triangles) and $x = 0.025$ (leftward-facing triangles) for (Sc$_{1 - x}$Lu$_x)_{3.1}$In (left axis) along with the average static internal field $H_{av}$ for $x = 0$ (right axis), obtained from the fits of ZF $\mu$SR measurements. The vertical axes are scaled with $\gamma_\mu$, the gyromagnetic ratio of a positive muon. (b) The longitudinal relaxation rate $1/T_1$, obtained from the LF $\mu$SR measurements.}
\label{muSR3}
\end{figure}

The Lu-doped samples show relaxing signal without precession at low temperatures, indicating a more random internal field, as compared with the undoped Sc$_{3.1}$In. In order to reproduce the observed line shape, the ZF time spectra of the Lu-doped samples have been analyzed by assuming the following functional form:

\begin{equation}
G(t) = A_1(1-p\sigma^2)e^{-\frac{1}{2}\sigma^2t^2} + \frac{A_1}{2}e^{(-\frac{t}{T_1})}+ A_{bg}
\label{GtF}
\end{equation}

\noindent where the first term represents the transverse relaxation, the second term is the longitudinal $1/T_1$ component, and the third term is a background signal from the sample holder. From independent measurements in weak transverse field at low temperatures, the values of the non-relaxing background signal from a silver sample holder $A_{bg}$ were estimated to be 0.36 and 0.55 for $x = 0.01$ and 0.025 samples, respectively. These values are consistent with the known background level from the cryostat and sample holder, and a rather small sample size. It is, however, not possible to eliminate the possibility that signal from a small paramagnetic volume in the specimen, persisting to $T = 0$ K, is included in the background signal. Due to difficulty in separating the effects of slow relaxation and partial volume fraction, the amplitude $A_1$ was fixed to be temperature-independent, allowing to extract the relaxation rate $\sigma$. A phenomenological "dip" parameter $p$ ($p = 1$) for the Kubo-Toyabe function was introduced, while smaller $p$ values would fit line shapes with a shallower dip, which are often observed in real materials, including the present case of Lu-doped systems. Although the fit is not perfect, as shown by the lines in Fig. \ref{muSR1}, the functional form of Eq. \ref{GtF} was used to compare the relaxation rates in different specimens without introducing additional free parameters. Fig. \ref{muSR3}(a) shows the temperature dependence of $\sigma$ in the two Lu-doped specimens. To compare the relaxation rates $\sigma$ with the static field measured in undoped Sc$_{3.1}$In, the spatially averaged value of the static local field was determined as:

\begin{equation}
H_{av} = V_M \frac{A_1\nu_1+ A_2\nu_2}{A_1+A_2} + \frac{(1-V_M)(\Delta_{KT1}+\Delta_{KT2})}{2\gamma_\mu}
\label{Hav}
\end{equation}

\noindent where $\gamma_\mu$ represents the gyromagnetic ratio of a positive muon and $\Delta_{KT}$ are the widths of the Kubo-Toyabe function for nuclear dipolar fields. We plot $H_{av}$ in Fig. \ref{muSR3}(a) with the relaxation rate (left axis) and the average field (right axis), scaled with $\gamma_\mu$. Since the static internal field is expected to be proportional to the local static spin polarization, Fig. \ref{muSR3}(a) demonstrates the development of the spatially-averaged magnetic order parameter which can be compared to the spontaneous magnetization $M_0$, shown in Fig. \ref{M0}.

Measurements of the spin-lattice relaxation rate $1/T_1$ were performed in an applied LF. Figure \ref{muSR3}(b) shows the temperature dependence of $1/T_1$ for (Sc$_{1-x}$Lu$_x$)$_{3.1}$In with $x = 0, 0.01, 0025$. There is no divergent behavior in Sc$_{3.1}$In, while the Lu-doped samples exhibit a peak in $1/T_1$ at the ordering temperature. In either case, the absolute values of $1/T_1$ are less than 0.1/$\mu$s, which implies that the relaxation rate measured in zero field (Fig. \ref{muSR3}(a)) is predominantly due to a static field, even at temperatures very close to the ordering temperature. In Fig. \ref{muSR3}(a), a finite relaxation rate/average field persists up to high temperatures for all the three systems. This is attributed to the nuclear dipolar field, as Sc has a very large nuclear moment. 

The absence of dynamic critical behavior and the gradual change of the volume fraction $V_M$, observed in undoped Sc$_{3.1}$In, indicates a first-order transition at the magnetic order. It is interesting to note that a weak "second order" feature is observed for Lu-doped samples. However, further experimental data are needed given the fact that the order parameter (Fig. \ref{muSR3}(a)) exhibits a non-linear dependence on the Curie temperature $T_C$, suggesting a remaining effect of first-order quantum evolution. Additionally, the difficulty in separating the effects of moment sizes and volume fraction at very small relaxation rates, as well as the uncertainty in estimating background level, prevent reliable determination of $V_M$ for the Lu-doped samples.  

\section{Discussion}

\begin{figure*}
\includegraphics[width=2\columnwidth]{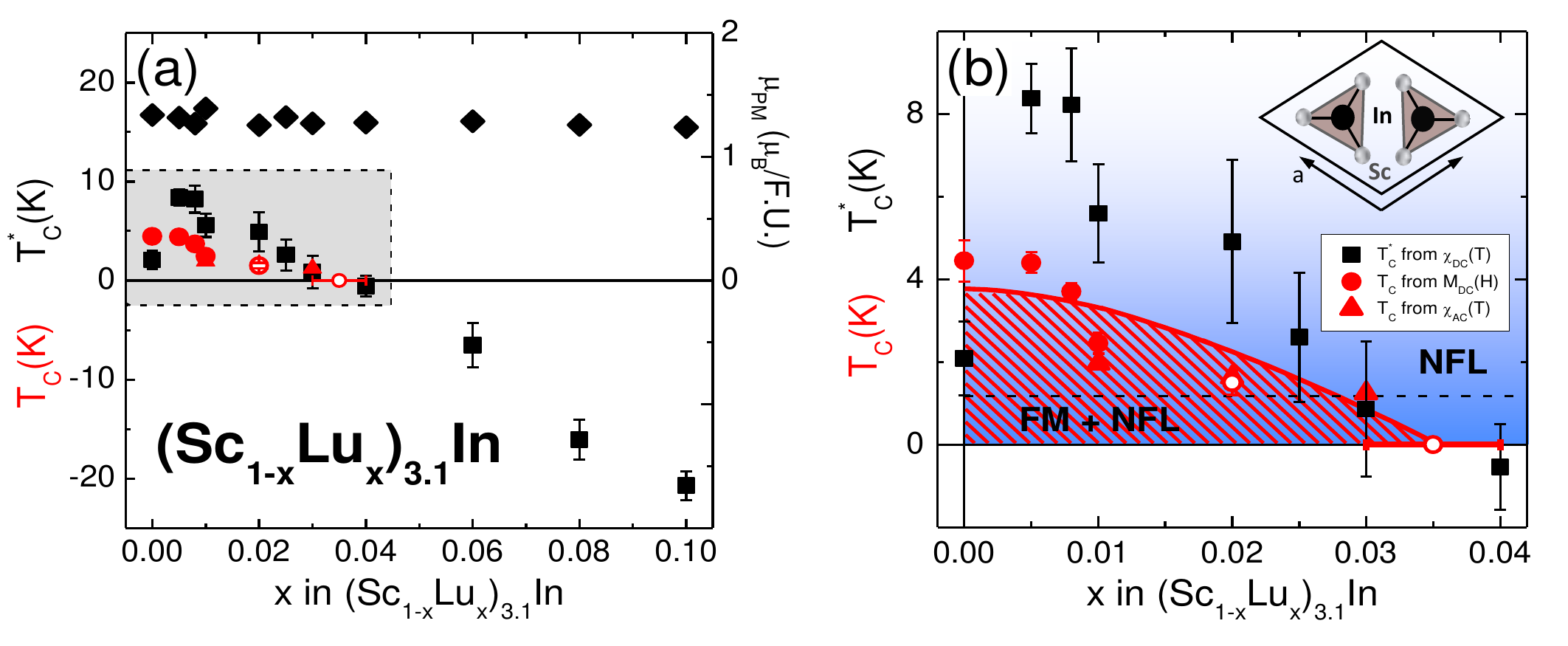}
\caption{(a) $T - x$ phase diagram for (Sc$_{1-x}$Lu$_x$)$_{3.1}$In for $0~\leq~x~\leq~0.10$, with the Weiss-like temperature $T^*_C$ (squares, left axis), the Curie temperature $T_C$ (circles, left axis) and paramagnetic moment $\mu_{PM}$ (diamonds, right axis). (b) Enlarged $T - x$ phase diagram for $x \leq 0.04$ [shaded area in (a)], with the \textit{ab} plane projection of the Sc$_{3.1}$In unit cell shown in the inset. The horizontal line at $T = 1.17$ K denotes the minimum experimental temperature, with the open symbols representing $T_C$ estimates extrapolated from accessible measurements.} 
\label{Parameters}
\end{figure*}

The paramagnetic moment $\mu_{PM}$ (diamonds, Fig. \ref{Parameters}(a)), determined from the Curie-Weiss\textit{-like} law, is nearly composition-independent $\mu_{PM} \sim 1.3 \mu_B/F.U.$ Moreover, the Weiss-\textit{like} temperature $T^*_C$ decreases nearly linearly with $x$ for $x \leq 0.10$, after an initial jump between $x = 0$ and 0.005 (squares, Fig. \ref{Parameters}(a)). Considering that Curie-Weiss\textit{-like} behavior in the itinerant scenario arises from the temperature-dependence of the amplitude of spin fluctuations \cite{mor}, this sudden increase in the corresponding $T^*_C$ signals enhanced spin fluctuations due to the disorder brought on by Lu doping. Between $x = 0.02$ and $x = 0.04$, $T^*_C$ changes sign in a continuous manner, suggesting the presence of a (second order) doping-induced QCP in this composition range near $x_c = 0.035~\pm~0.005$. Moreover, $T_C$ determined either from $\chi'(T)$ or $M(H)$ data (Figs. \ref{MT}(c)-\ref{AN}) moves down in temperature below 1.17 K for the doping amounts above 0.03, indicating that the QCP is close to this composition.

The determination of the critical composition $x_c$ at the QCP requires consistency between the $M(T)$ and $\chi'(T)$ data, the critical scaling analysis of the $M(H)$ measurements as well as the $\mu$SR results. Indeed, the critical composition $x_c = 0.035 \pm 0.005$ is determined from (i) the $T_C$ (circles and triangles) and $T^*_C$ (squares) values (Fig. \ref{Parameters}(b)) approaching 0 K at the QCP and (ii) the critical scaling rendering the Arrott-Noakes plots $M^{1/\beta}$ \textit{vs.} $(H/M)^{1/\gamma}$ as parallel isotherms, equally spaced in $t$ (Figs. \ref{Sc3In}(c) and \ref{AN}(a-c)). Moreover, the $\mu$SR results confirm the development of static magnetic order with a nearly full volume fraction at low temperatures, and diminishing moment size as a function of decreasing ordering temperature. Moreover, the continuous variation of $T^*_C$ and $T_C$ with $x$ and the $\mu$SR evidence for a second order phase transition in the Lu doped samples are also evidence for the QPT induced by Lu doping.

Doping in Sc$_{3.1}$In reveals intriguing traits associated with quantum criticality in general, and with weak IFM systems in particular: the paramagnetic moment $\mu_{PM}$ is surprisingly large in (Sc$_{1-x}$Lu$_x$)$_{3.1}$In, and nearly independent of $x$, even as the system goes through the QPT at $x_c = 0.035 \pm 0.005$. Not surprisingly then, the critical exponent $\beta$ is unchanged through the ferromagnetic state, although its value $\beta = 0.26~\pm~0.05$ is smaller than that in any other known quantum critical system. The minute critical composition and small $\beta$ value, together with the jump in $T^*_C$ as $x~>~0$ (Fig. \ref{Parameters}(b), squares) point to a weak IFM ground state, easily perturbed by doping. This may seem unusual in light of the stark differences between Sc$_{3.1}$In and the related IFM system ZrZn$_2$ \cite{sok}, well described by mean-field theory, or the similarities with the extraordinary critical scaling in the HF FM URu$_{2-x}$Re$_x$Si$_2$ \cite{but}, close to these systems' respective doping-induced QCPs. However, these observations may be reconciled from crystallographic and electronic properties considerations: as a nearly 1D structure is formed by bipyramidal Sc-In chains (inset of Fig. \ref{Xrays}), the reduced dimensionality in Sc$_{3.1}$In renders it more similar to the layered (2D) URu$_2$Si$_2$ compound than the cubic (3D) ZrZn$_2$. It appears that the NFL behavior in the ferromagnetic state may also be correlated with the non-mean-field scaling, and, more importantly, that this correlation is independent of the presence of hybridized $f$-electrons. Consequently, the universality of the quantum critical behavior common to the former two compounds may be ascribed to spin fluctuations, associated with reduced crystallographic dimensionality. More IFM systems are needed to probe this universality. Equally important is the synthesis of single crystals of Sc$_{3.1}$In, which would enable further characterization of the implications of dimensionality on the QCP, as well as to probe the potential NFL behavior at the QCP and in the ferromagnetic state. These experiments are currently underway.

\section{Acknowledgments}

We thank L. L. Zhao for help with figure edits, P. C. Canfield for providing some of the rare earth metals, and Q. Si and M. C. Aronson for useful discussions. The work at Rice was supported by NSF DMR 0847681. Measurements of the AC magnetic susceptibility were performed at UCSD with support from the National Science Foundation under Grant No. DMR-1206553. Work at Columbia and TRIUMF (L.L., B.F. and Y. J. U.) is supported by NSF grants DMR-1105961 and OISE-0968226 (PIRE), REIMEI project from JAEA, Japan, and the Friends of Todai Inc. Foundation. Work at IOPCAS was supported by NSF and MOST of China through research projects.

\end{document}